# Synthesis and post-annealing effects of alkaline-metal-ethylenediamine-intercalated superconductors $A_x(C_2H_8N_2)_yFe_{2-z}Se_2$ ($A$ = Li, Na) with $T_c$ = 45 K

Takashi Noji, Takehiro Hatakeda, Shohei Hosono, Takayuki Kawamata, Masatsune Kato, Yoji Koike

*Department of Applied Physics, Graduate School of Engineering, Tohoku University,*

*6-6-05 Aoba, Aramaki, Aoba-ku, Sendai 980-8579, Japan*

New iron-based intercalation superconductors $A_x(C_2H_8N_2)_yFe_{2-z}Se_2$ ($A$ = Li, Na) with $T_c$ = 45 K have successfully been synthesized via intercalation of dissolved alkaline metal in ethylenediamine. The *c*-axis lengths of $A_x(C_2H_8N_2)_yFe_{2-z}Se_2$ ($A$ = Li, Na) are 20.74(7) Å and 21.9(1) Å, respectively, and are about 50 % larger than that of $K_xFe_2Se_2$, indicating that not only alkaline metal but also ethylenediamine is intercalated between the Se–Se layers of FeSe. It seems that the high-$T_c$ of $A_x(C_2H_8N_2)_yFe_{2-z}Se_2$ ($A$ = Li, Na) is caused by the possible two-dimensional electronic structure due to the large *c*-axis length. Through the post-annealing in an evacuated glass tube, it has been found that $T_c$ decreases with increasing post-annealing temperature and that deintercalation of EDA from the as-intercalated sample takes place at low temperatures below 250℃.

## 1. Introduction

The superconducting transition temperature, $T_c$, of the iron-based chalcogenide superconductor FeSe is only 8 K [1], but dramatically increases up to ~37 K by the application of high pressure [2]. The successful synthesis of potassium-intercalated $K_xFe_{2-y}Se_2$ with $T_c$ ~31 K at ambient pressure was an exciting breakthrough [3]. Recently, it has been reported that the intercalation into FeSe of alkaline and alkaline-earth metals in liquid ammonia yields a variety of compounds with significantly enhanced $T_c$'s of 40 - 46 K [4 - 6]. The c-axis length of $Li_{0.9}Fe_2Se_2(NH_3)_{0.5}$ with $T_c$ = 44 K is 16.518 Å [5] and larger than 14.0367 Å in $K_xFe_2Se_2$ [3], indicating that not only lithium but also ammonia is intercalated. Moreover, it has been reported that lithium- and pyridine-intercalated $Li_x(C_5H_5N)_yFe_{2-z}Se_2$ exhibits superconductivity with $T_c$ = 45 K and that post-annealing of the intercalated sample drastically expands the $c$-axis length from 16.0549 Å to 23.09648 Å and increases the superconducting shielding volume fraction [7].

Here, we report on the successful synthesis of new superconductors $A_x(C_2H_8N_2)_yFe_{2-z}Se_2$ ($A$ = Li, Na) with $T_c$ = 45 K via intercalation of dissolved alkaline metal in ethylenediamine (EDA), $C_2H_8N_2$. Post-annealing effects on the crystal structure and superconductivity are also discussed.

## 2. Experimental

Polycrystalline samples of FeSe were prepared by the solid-state reaction method. Starting materials were powders of Fe and Se, which were weighted stoichiometrically, mixed thoroughly and pressed into pellets. The pellets were sealed in an evacuated quartz tube and heated at 800℃ for 40 h. The obtained pellets of FeSe were pulverized into powder to use for the intercalation. Dissolved alkaline metal in EDA was intercalated into the powdery FeSe as follows. An appropriate amount of the powdery FeSe was placed in a beaker filled with 0.2 M solution of pure lithium or sodium metal dissolved in EDA. All the process was performed in an argon-filled glove box. Post-annealing of as-intercalated samples was carried out at 100 - 500℃ for 60 h in an evacuated glass tube. Both FeSe and the intercalated samples were characterized by powder x-ray diffraction using Cu $K_\alpha$ radiation. For the intercalated samples, an airtight sample holder was used. In order to observe the superconducting transition, the magnetic susceptibility, $\chi$, was measured using a superconducting quantum interference device (SQUID) magnetometer. Measurements of the electrical resistivity, $\rho$, were also carried out by the standard dc four-probe method. For the $\rho$ measurements, as-intercalated powdery samples were pressed into pellets. Then, the pellets were sintered at 200℃ for 20 h in an evacuated glass tube. Thermogravimetric (TG) measurements were performed in flowing gas of argon, using a commercial analyzer (SII Nano Technology Inc., EXSTAR DSC7020).

## 3. Results and discussion

Figure 1 shows powder x-ray diffraction patterns of as-intercalated samples of $A_x(C_2H_8N_2)_yFe_{2-z}Se_2$ ($A$ = Li, Na). The broad peak around $2\theta = 20°$ is due to the airtight sample holder. Although there are unknown peaks, most of sharp Bragg peaks are due to the intercalation compound of $A_x(C_2H_8N_2)_yFe_{2-z}Se_2$ ($A$ = Li, Na) and the host compound of FeSe, so that they are able to be indexed based on the $ThCr_2Si_2$-type and PbO-type structures, respectively. Therefore, it is found that alkaline metal and EDA are partially intercalated into FeSe, while there remains a non-intercalated region of FeSe in the samples. The lattice constants of $Li_x(C_2H_8N_2)_yFe_{2-z}Se_2$ are calculated to be $a$ = 3.458(6) Å and $c$ = 20.74(7) Å. The c-axis length of $Na_x(C_2H_8N_2)_yFe_{2-z}Se_2$ is 21.9(1) Å. Since the unit cell of $A_x(C_2H_8N_2)_yFe_{2-z}Se_2$ ($A$ = Li, Na) includes two Fe layers, the distance between the neighboring Fe layers is 10.37(4) Å and 10.95(5) Å, respectively, and much larger than 5.515(1) Å of FeSe. Taking into account our previous results that the intercalation of only lithium into Fe(Se,Te) has neither effect on the superconductivity nor crystal structure [8], it is concluded that not only lithium or sodium but also EDA has been intercalated between the Se-Se layers of FeSe.

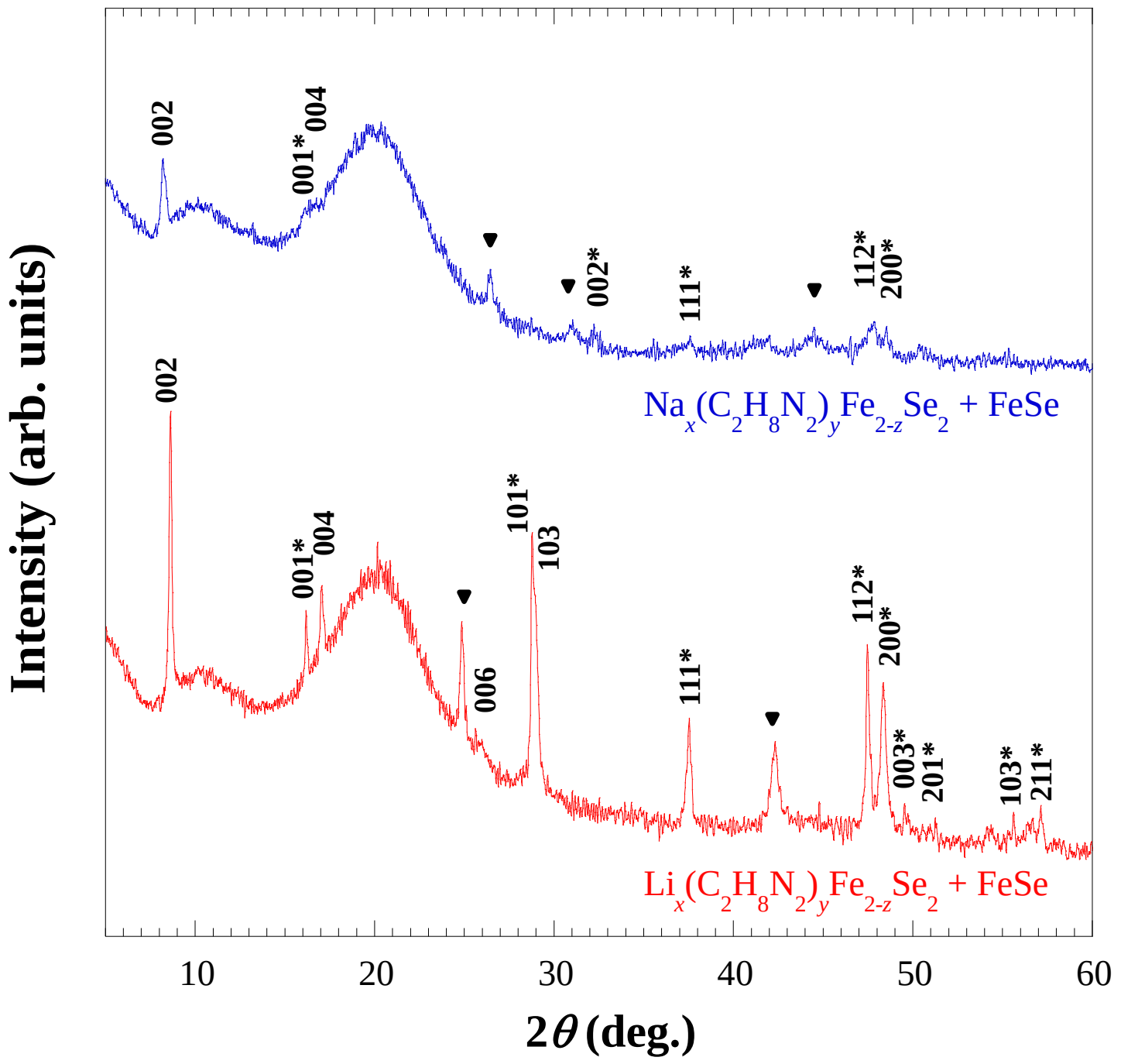

Fig. 1. Powder x-ray diffraction patterns of as-intercalated samples consisting of $A_x(C_2H_8N_2)_yFe_{2-z}Se_2$ ($A$ = Li, Na) and FeSe using Cu K$_\alpha$ radiation. Indexes without and with asterisk are based on the $ThCr_2Si_2$-type and PbO-type structures, respectively. Peaks marked by ▼ are unknown.

Figure 2 shows the temperature dependence of $\chi$ in a magnetic field of 10 Oe on zero-field cooling (ZFC) and on field cooling (FC) for as-intercalated powdery samples. The first $T_c$ is observed at 45 K and the second $T_c$ is at 8 K. Taking into account the powder x-ray diffraction results, it is concluded that the first is due to bulk superconductivity of $A_x(C_2H_8N_2)_yFe_{2-z}Se_2$ ($A$ = Li, Na), while the second is due to that of the non-intercalated region of FeSe. The inset shows the temperature dependence of $\rho$ for the sintered (200℃, 20 h) pellet sample. Although the as-intercalated sample pelletized simply at room temperature did not show zero resistivity, it is found that the resistivity of the sintered pellet sample starts to decrease at 43 K with decreasing temperature and reaches zero at 18 K. Since the intercalation of only lithium does not increase $T_c$ [8], it seems that the high-$T_c$ of $A_x(C_2H_8N_2)_yFe_{2-z}Se_2$ ($A$ = Li, Na) is caused by the possible two-dimensional electronic structure due to the large c-axis length.

Figure 3 shows the TG curve on heating up to 900℃ at the rate of 1℃/min for the as-intercalated powdery sample. Three steps of mass loss are observed; 2.9 % loss below 100℃, 11.3 % loss between 100 and 200℃, and large loss above 450℃.

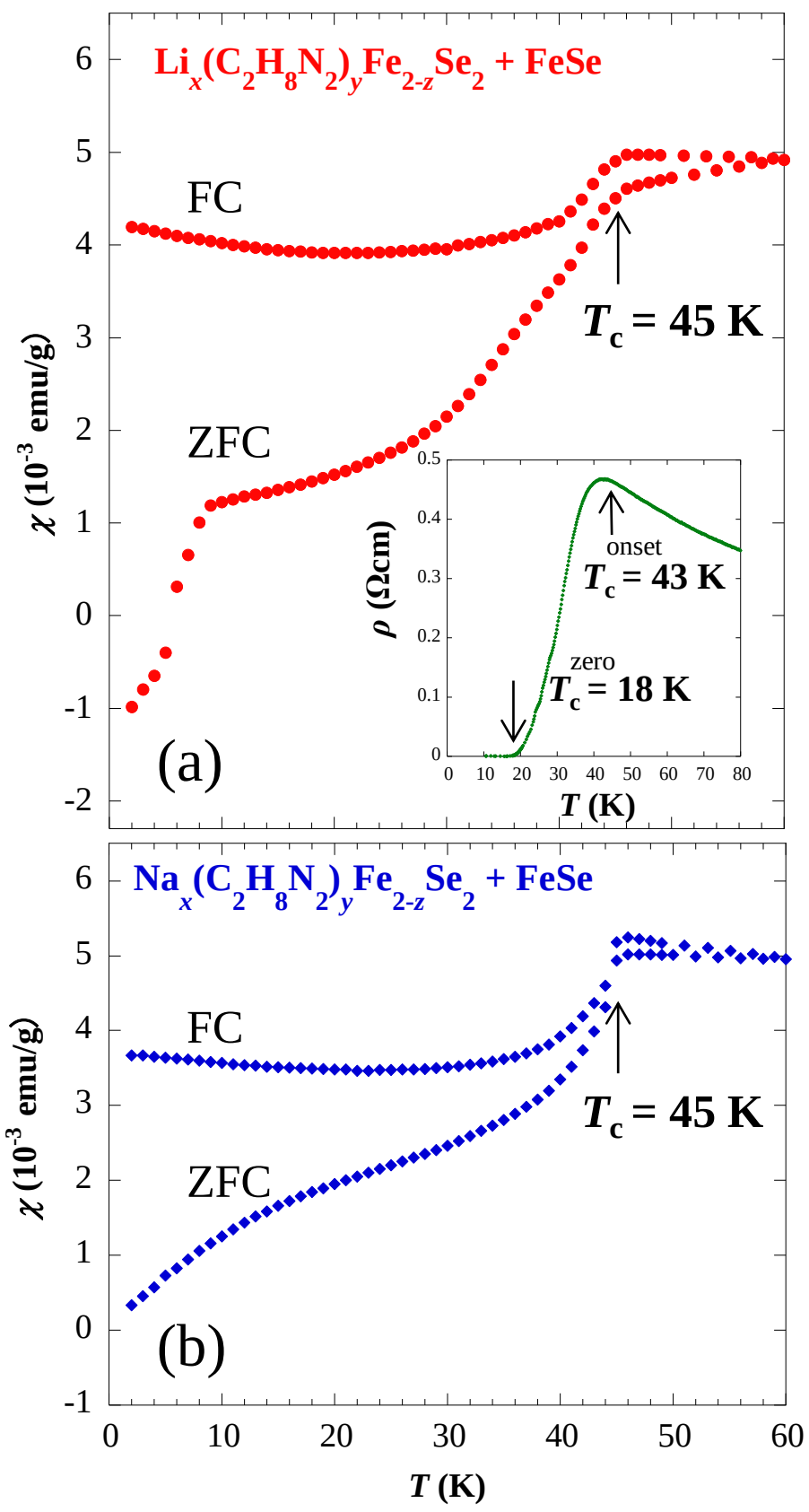

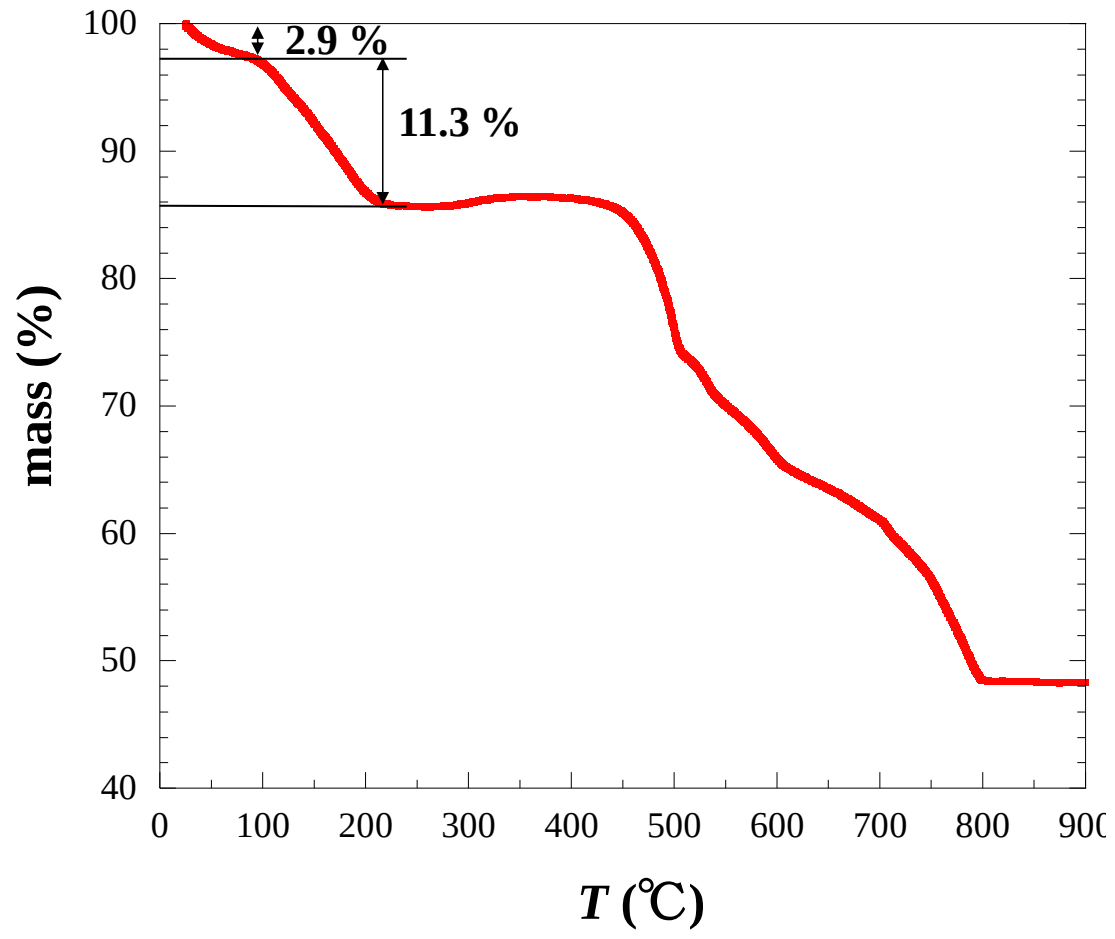

Fig. 3. Thermogravimetric (TG) curve on heating at the rate of 1℃/min for the as-intercalated powdery sample consisting of $Li_x(C_2H_8N_2)_yFe_{2-z}Se_2$ and FeSe.

Fig. 2. Temperature dependence of the magnetic susceptibility, $\chi$, in a magnetic field of 10 Oe on zero-field cooling (ZFC) and field cooling (FC) for as-intercalated powdery samples consisting of $A_x(C_2H_8N_2)_yFe_{2-z}Se_2$ ((a)$A$ = Li, (b)$A$ = Na) and FeSe. The inset shows the temperature dependence of the electrical resistivity, $\rho$, for the sintered (200℃, 20 h) pellet sample consisting of $Li_x(C_2H_8N_2)_yFe_{2-z}Se_2$ and FeSe.

Figure 4 shows powder x-ray diffraction patterns of the as-intercalated sample and samples post-annealed at various temperatures for 60 h. For samples post-annealed below 200℃, diffraction peaks of $Li_x(C_2H_8N_2)_yFe_{2-z}Se_2$ remain. For samples post-annealed at 250℃ and 300℃, only peaks due to FeSe are observed, and these peaks disappear for the sample post-annealed at 500℃. Accordingly, the first step of the TG curve below 100℃ may be due to desorption of EDA on the surface of grains. The second step between 100 and 220℃ is due to deintercalation of EDA. The third step above 500℃ may be due to unknown products including iron.

Figure 5 shows the temperature dependence of $\chi$ in a magnetic field of 10 Oe on ZFC and FC for as-intercalated and post-annealed (100 – 500℃, 60 h) powdery samples. The $T_c$ of $Li_x(C_2H_8N_2)_yFe_{2-z}Se_2$ decreases with increasing post-annealing temperature. For samples post-annealed at 250℃ and 300℃, only the superconducting transition of FeSe is observed, and the transition also disappears for the sample post-annealed at 500℃. These results are in good correspondence to the results of x-ray diffraction.

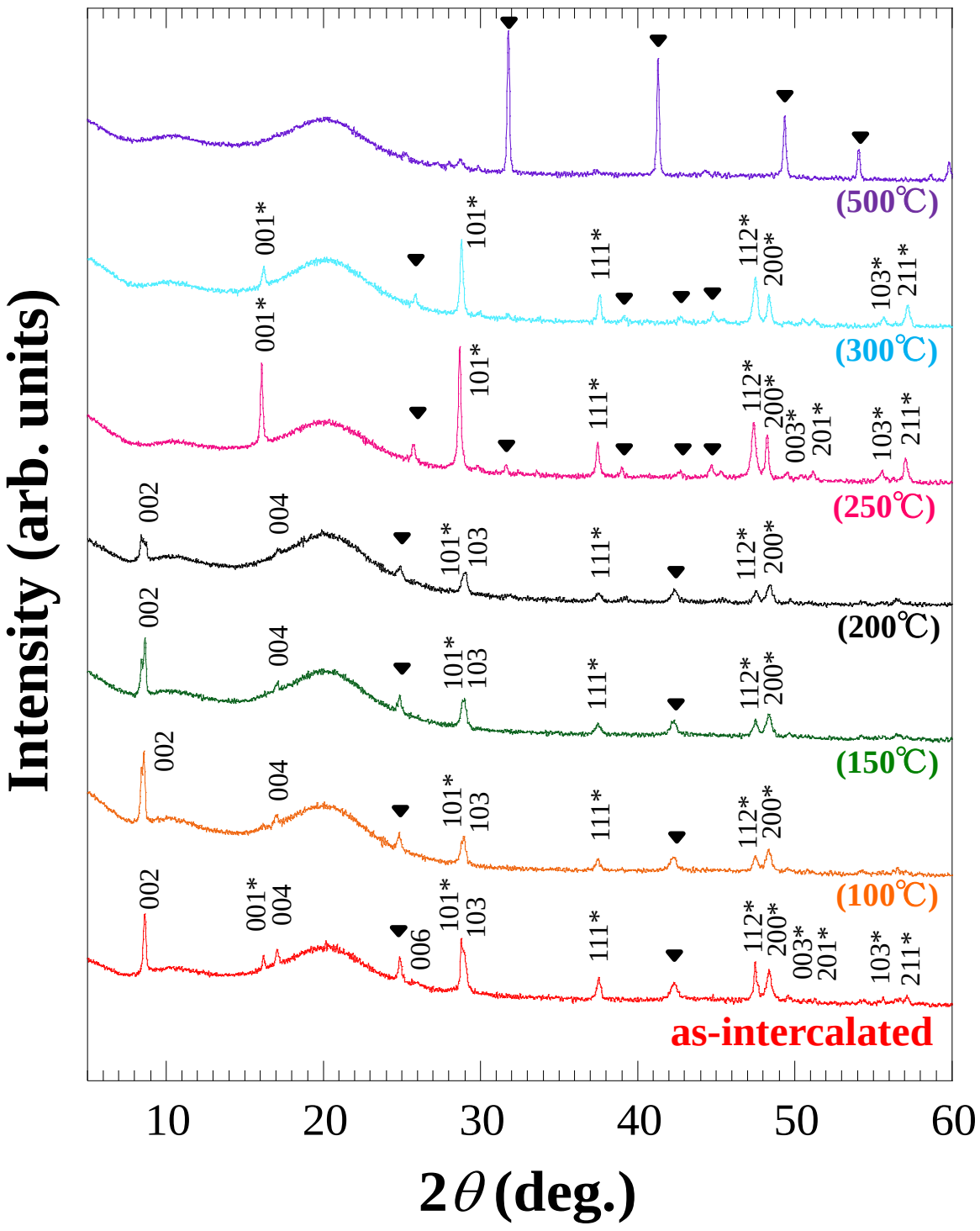

Fig. 4. Powder x-ray diffraction patterns of as-intercalated and post-annealed (100 - 500℃, 60 h) samples using Cu $K_\alpha$ radiation. Indexes without and with asterisk are based on the $ThCr_2Si_2$-type and PbO-type structures, respectively. Peaks marked by▼ are unknown.

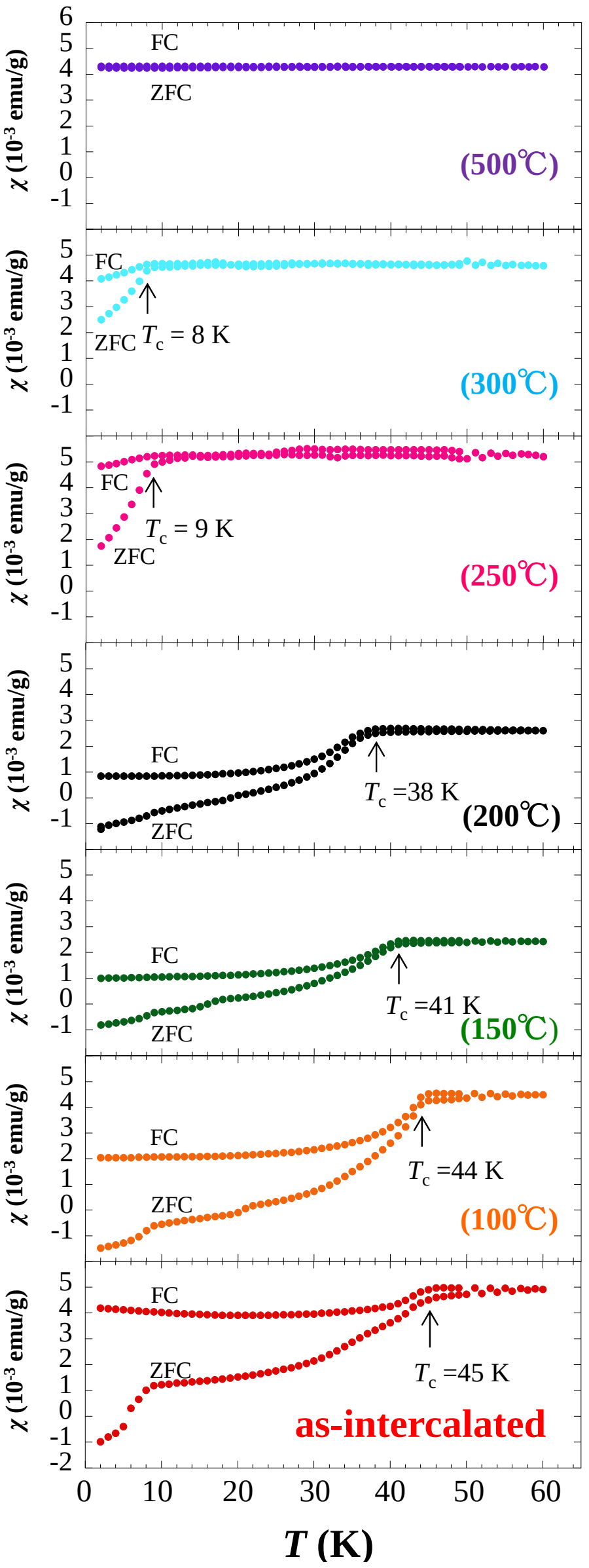

Fig. 5. Temperature dependence of the magnetic susceptibility, $\chi$, in a magnetic field of 10 Oe on zero-field cooling (ZFC) and field cooling (FC) for as-intercalated and post-annealed (100 - 500℃, 60 h) powdery samples consisting of $Li_x(C_2H_8N_2)_yFe_{2-z}Se_2$ and FeSe.

Finally, Fig. 6 shows the maximum $T_c$ 's of FeSe-based superconductors obtained so far at ambient pressure as a function of the distance between the neighboring Fe layers. Interestingly, $T_c$ increases monotonically with increasing such distance and is saturated at about 45 K above 9 Å.

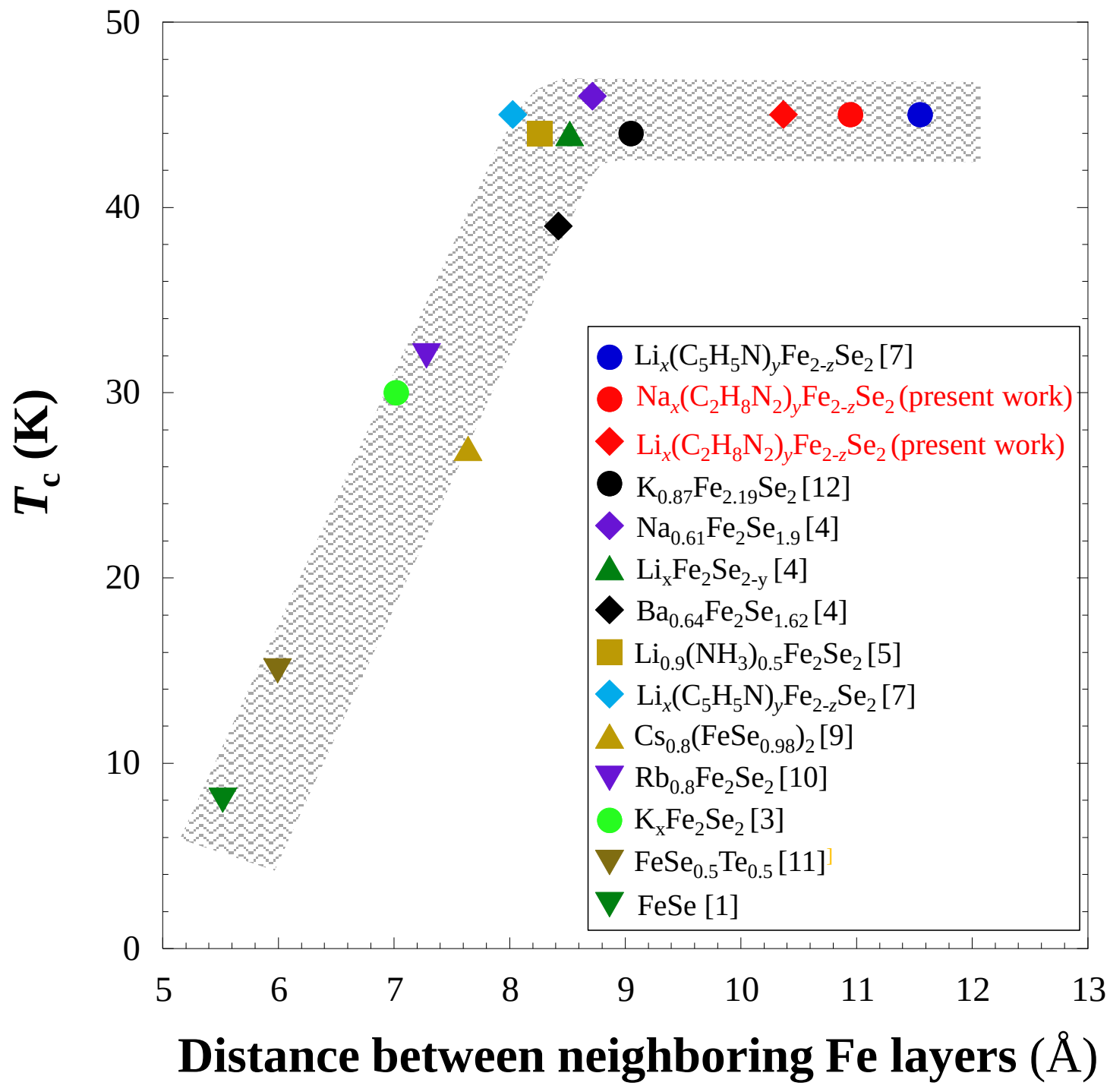

Fig. 6. Relation between the maximum $T_c$ and the distance of the neighbouring Fe layers of FeSe-based superconductors.

## 4. Summary

We have succeeded in synthesizing new intercalation compounds $A_x(C_2H_8N_2)_yFe_{2-z}Se_2$ ($A$ = Li, Na) via intercalation of dissolved alkaline metal in EDA. Although the samples include non-intercalated regions of FeSe, the $c$-axes of $A_x(C_2H_8N_2)_yFe_{2-z}Se_2$ ($A$ = Li, Na) have drastically expanded to be 20.74(7) and 21.9(1) Å, respectively, indicating that not only alkaline metal but also EDA is intercalated between the Se-Se layers of FeSe. Bulk superconductivity has been observed below 45 K in the $\chi$ measurements of as-intercalated samples of $A_x(C_2H_8N_2)_yFe_{2-z}Se_2$ ($A$ = Li, Na). Moreover, zero resistivity has been observed below 18 K for the sintered pellet sample of $Li_x(C_2H_8N_2)_yFe_{2-z}Se_2$. It seems that the high-$T_c$ of $A_x(C_2H_8N_2)_yFe_{2-z}Se_2$ ($A$ = Li, Na) is caused by the possible two-dimensional electronic structure due to the large c-axis length. Through the post-annealing in an evacuated glass tube, it has been found that $T_c$ decreases with increasing post-annealing temperature and that deintercalation of EDA from the as-intercalated sample takes place at low temperatures below 250℃ and that FeSe remains at temperatures around 400℃. It has also turned out that the maximum $T_c$ 's of FeSe-based superconductors rise with increasing distance between the neighboring Fe layers and are saturated at about 45 K above 9 Å.

## References

[1] F.-C. Hsu, J.-Y. Luo, K.-W. Yeh, T.-K. Chen, T.-W. Huang, P. M. Wu, Y.-C. Lee, Y.-L. Huang, Y.-Y. Chu, D.-C. Yan, and M.-K. Wu, Proc. Natl. Acad. Sci. U.S.A **105** (2008) 14262 - 14264.

[2] S. Margadonna, Y. takabayashi, Y. Ohishi, Y. Mizuguchi, Y. Takano, T. Kagayama. T. Nakagawa, M. Takata, and K. Prassides, Phys. Rev. B **80** (2009) 064506/1 - 6.

[3] J. Guo, S. Jin, G. Wang, S. Wang, K. Zhu, T. Zhou, M. He, and X. Chen, Phys. Rev. B **82** (2010) 180520(R)/1 - 4.

[4] T. P. Ying, X. L. Chen, G. Wang, S. F. Jin, T. T. Zhou, X. F. Lai, H. Zhang, and W. Y. Wang, Sci. Rep. **2** (2012) 426/1 - 7.

[5] E.-W. Scheidt, V. R. Hathwar, D. Schmitz, A. Dunbar, W. Scherer, F. Mayr, V. Tsurkan, J. Deisenhofer, and A.Loidl, Eur. Phys. J. B **85** (2012) 279/1 - 5.

[6] M. Burrard-Lucas, D. G. Free, S. J. Sedlmaier, J. D. Wright, S. J. Cassidy, Y. Hara, A. J. Corkett, T. Lancaster, P. J. Baker, S. J. Blundell, and S. J. Clarke, Nat. Mater. **12** (2013) 15 - 19.

[7] A. Krzton-Maziopa, E. V. Pomjakushina, V. Y. Pomjakushin, F. Rohr, A. Schilling, and K. Conder, J. Phys.: Condens. Matter **24** (2012) 382202/1 - 6.

[8] H. Abe, T. Noji, M. Kato, and Y. Koike, Physica C **470** (2010) S487 - S488.

[9] A. Krzton-Maziopa, Z. Shermadini, E. Pomjakushina, V. Pomjakushin, M. Bendele, A. Amato, R. Khasanov, H. Luetkens, and K. Conder, J. Phys.: Condens. Matter **23** (2011) 052203/1 - 4.

[10] A. F. Wang, J. J. Ying, Y. J. Yan, R. H. Liu, X. G. Luo, Z. Y. Li, X. F. Wang, M. Zhang, G. J. Ye, P. Cheng, Z. J. Xiang, and X. H. Chen, Phys. Rev. B **83** (2011) 060512(R)/1 - 4.

[11] K.-W. Yeh, T.-W. Huang, Y.-L. Huang, T.-K. Chen, F.-C. Hsu, P. M. Wu, Y.-C. Lee, Y.-Y. Chu, C.-L. Chen, J.-Y. Luo, D.-C. Yan, and M.-K. Wu: Europhys. Lett. **84** (2008) 37002/1 - 4.

[12] A. –M. Zhang, T. –L. Xia, K. Liu, W. Tong, Z. –R. Yang, and Q. –M. Zhang, Sci. Rep. **3** (2013) 1216/1 - 5.